# Georg de Buquoy – Founder of Mathematical Economy with South Bohemian Roots

Dalibor Štys[1], Miroslava Vlačihová

**Abstract.** Georg de Buquoy, Lord de Vaux, lived in Nové Hrady, Prague and Červený Hrádek for most of his productive life. From his extensive scientific contributions, both theoretical and experimental, we expand here the discussion of his contributions to mathematical economy. He is mainly celebrated as the first persons to define correctly net yield and describe method for its optimization, which was considered "strikingly modern" still in 1950´ [1]. Buquoy´s program was "systematic overview of all theorems which affect maintenance and increase of national wealth" [2] for which he correctly defined and mathematically expressed many economic terms [3].

The most striking feature of Buquoy´s writing is that he was also a very influential economic practicioner. He governed one of the wealthiest possessions in Bohemia, and perhaps in Austria, of his time. Thus his economical thinking expands from the "political part", which is economy in modern sense, to "…sources of national wealth, or the technical part of national economy …". The complexity of Buquoy´s view has little match in modern literature namely because the extent of data sources is hardly available in modern times.

## 1 Introduction

Georg Frances August de Lonqueval, Comte de Buquoy, Lord de Vaux, (1781-1851) was born in Brussels at September 17$^{th}$, 1781. In 1795 died his father and his uncle, Bohemian majority lord Johann Nepomuk von Buquoy, adopted him. He studied privately in Prague where he passed public Gymnasium examinations and later in the Theresian Knight Academy in Vienna, philosophy and law. However, his prime subjects obviously were mathematics and physics. As he wrote at several places in introductions to his work, he "studied day and night… the subjects of mathematics and physics … primarily the works of French authors.

In his biography are highly interesting his voyages to Europe. The first one took place in 1803-1805, a time of large political instability in Europe. The reports from this voyage are can be extracted from his diary which is highly comprehensive. It is an interesting testimony of his time (Mann, unpublished edition of the diary of J. von Buquoy). The voyage obviously gave him a lot of inspiration, namely in the industrial development of his possession, he himself later found it "disattaractng him from his main interest in mathematics". Second trip was to Paris in 1815 where he at the Academy of Science presented his theory of virtual speeds. His lecture was recognized by physicists of his time, namely Laplace. But when Buquoy tried to present solution of wave problems which obviously preoccupied scientific discussions of that time and which was long time ago known to Gerstner, he faced many obstacles. Finally, he broke relations to French scientific community and focused on his German and Scandinavian colleagues.

Georg de Buquoy lived in Nové Hrady, Prague and Červený Hrádek for most of his productive life. He was active in many fields of science starting from mechanics, through chemistry, thermodynamics, optics up to economics, political theory and theory of speech and its interpretation, in modern terms, cognitive science. For example the term Biology he used in 1817 [4] for the first time on the Czech ground. He was also known as extremely successful enterpreneur, he invented among other the special sort of dark heavy glass, the hyalit. His original steam engine was in 1812 the first built at Czech ground. To his industrial success contributed highly his marriage with Gabriela von Rottenhan, the heir of a large possession in northern Bohemia which included coal and ore mines. Buquoy was friend of the biggest personality of German culture in his times, Goethe, and many other scientists and politicians. Most of these people he met in Karlovy Vary a place relatively close to his possessions.

---

[1] Institute of Physical Biology, University of South Bohemia, Zámek 136, 373 33 Nové Hrady.

At the end of his life Buquoy participated in the Prague revolution 1848, the uprisal against the Habsburg empery. Buquoy, who considered Austria "land of obscurantism", participated in the organizational structures and donated money. He was set to house arrest for his alleged concernment on Czech crown. He died in Prague at April 19[th], 1851.

## 2 Buquoys scientific achievements

### 2.1 Buquoys mechanics – theory of virtual speeds

Among his achievements are mostly cited his works in mechanics [5,6]. The latest direst citation known to us comes from Galperin [7] in 2008. Buquoy´s work is a breakthrough generalisation of Newtons law where the notoriously known equation

$$F(t,x(t),v(t)) = mx'', v(t) = x'(t), x''(t) = v'(t) = a(t), x(0) = x_0, v(0) = v_0, t \geq 0 \tag{1}$$

Where $F(t,x(t),v(t))$ is mechanical force, $x(t)$ trajectory, $v(t)$ velocity, $a(t)$ acceleration and $t$ time. Buquoy introduced more general form

$$d(mv) = mdv + vdm = F(t,c(t),v(t))dt, t \geq 0, dt > 0 \tag{2}$$

The later equation has been best recognized as the equation for rocket movement, although it was never derived with this purpose in Buquoys time. However, the fact that any body undergoing real movement loses the mass by burning fuel or, as in one Buquoy´s examples, increases the mass by mud adhering to carriage wheel, makes the equation (2) the ultimately correct at any instant.

However, the ultimate generalization of the equation (2) was left completely unattended in a strange philosophical half poem- half prose book book "*Ideal laudation of empirically intercepted life of Nature*" [8] where at the bottom remark at side 83 and 84 Buquoy expands

$$dv = \frac{F(t)}{m(t)} dt \tag{3}$$

And concludes that (in modern notation)

$$x = \int dt \int \frac{F(t)}{m(t)} dt = T(t) \tag{4}$$

And comments that from this equation my be read in two ways – time is a function of trajectory through the $t = \varphi(x)$ and space is a function of time through the function $x = T(t)$. Later in the text Buquoy expands one specific example when $\frac{F(t)}{m(t)} = k$ but the generality of the equation (4) remains untouched. As the whole concept of equivalence of moment/energy and mass and its generalized relation to time-space coordinates introduced *in priciple* in equations (1-4). The reader should refer for example to the article [9] for further (independent) specific expansion on this subject.

### 2.2 Buquoys economic thinking

Buquoy´s program was "systematic overview of all theorems which affect maintenance and increase of national wealth" [2] for which he correctly defined and mathematically expressed many economic terms [3]. He aims to achieve these goals in "purely technical" as well as in "political" fields. His main economical work is the "Theory of national economy [2], but it should be said that technical and political economical themes preoccupied his publications between 1827 and 1838 when he published more than 30 articles. These works were, to our knowledge, never fully analysed and surprises such as eq. (3) and (4) mentioned may be expected. Form the fist point of view, these late articles do no contain any new mathematical description and re not of interest for this article.

Buquoy is mainly celebrated as the first persons to define correctly net yield and describe method for its optimization, which was considered "strikingly modern" still in 1950´ [1]. The equation is

$$n = y - Y = f(x) - F(x) \tag{5}$$

where

$$y = f(x) \tag{6}$$

is the yield curve and

$$Y = F(x) \tag{7}$$

is the cost curve. To find how deep the farmer should plough the field, maximum has to be found.

It would be quite unjustly [3] to reduce Buquoy´s mathematical economy to net yield. He also correctly mathematised the relation between demand and market price, included the notion of competition. This is considered the first application of function in economic literature [1, 10]. Buquoy also mathematised the concept of natural price in terms of "real and nominal price of an item". The price is estimated by (a) the land rent $L$, (b) the running costs $u$, (c) interest from running costs $p$, and (d) interests from founding capital, i.e. fixed capital that needs to be maintained $F$.. So the natural value $W_1$ is for example for the flax producer the following: he should receive in the first instant coverage for the land rent $L$, yield which covers the necessary running costs $u_0$, and costs covering the yield form the fixed capital (i.e. capital to which is assumed infinite lifetime) $F_0$

$$W_1 = L + F_0 \frac{p}{100} + u_0 \left(1 + \frac{p}{100}\right) = L + \frac{F_0 p + u_0(100+p)}{100} \tag{8}$$

The natural price, i.e the price which the producer himself must receive, includes also the running costs of spinning $u_1$ and yield from the fixed capital of the spinner $u_1$

$$W_2 = \left(L + \frac{F_0 p + u_0(100+p)}{100} + u_1\right)\left(1 + \frac{p}{100}\right) + F_1 \frac{p}{100} \tag{9}$$

In this way, one may expand until the natural price for the end product where we get

$$W_n = L\left(1 + \frac{p}{100}\right)^n + \left(\frac{F_0 p + u_0(100+p)}{100}\right)\left(1 + \frac{p}{100}\right)^n + \left(\frac{F_1 p + u_1(100+p)}{100}\right)\left(1 + \frac{p}{100}\right)^{n-1} + \left(\frac{F_2 p + u_2(100+p)}{100}\right)\left(1 + \frac{p}{100}\right)^{n-2} + \ldots + \left(\frac{F_n p + u_n(100+p)}{100}\right) \tag{10}$$

By each new production step in the production chain the natural price inceases by multiplication by the factor $\left(1 + \frac{p}{100}\right)$. As the main thing – concludes Buquoy – is the fact that the costs of addition of each new production step the investors do not anticipate the costs of labour division. So the seeming reduction of costs has to be re-considered in this context.

Buquoy also devised equations which allow the entrepreneur to assess the feasibility of introduction of innovation at the market and also the influence of the competition. He defines the market price as

$$Y = \varphi\left(\overset{<}{y}, \overset{>}{x}\right) \tag{11}$$

Where $x$ is the competition at the sales side and $y$ is the demand. The symbols $f\left(\overset{<a>}{x}\right)$ mean "a value which depends on x in the way that to a certain value growths and from it decreases".

The most striking feature of Buquoy´s writing is that he was also a very influential economic practicioner. He governed one of the wealthiest private possessions in Bohemia, and perhaps in Austria, of his time. Thus his economical thinking expands from the "political part", which is economy in modern sense, to "…sources of national wealth, or the technical part of national economy …". The complexity of Buquoy´s view, namely in the technical part, has little match in modern literature. Namely because the extent of concrete and reliable data is hardly available in modern times.

## 2.3 Buquoys contribution to other branches of science

This approach well illustrates Buquoy´s approach to scientific thinking, the desire for generality of utilised mathematical formulae and the desire to utilise them for description of real world phenomena. In many cases, the experiments or observations described were of his own or repeated by himself in his own workshops or laboratories. Buquoys scientific methods were extensively discussed in the book "Ideal laudation of empirically intercepted life of Nature" [8].Here he writes (page xxxii) "The unavoidable utilization of mathematics comes mainly from dominant spatial spread of the inorganic (better sub-organic) *(form of matter)*. But what maters the laws of other phenomena, the qualitative laws of spatial *(mechanical)* phenomena have their analogies in laws of life. Mediating utilization of mathematics is here only allowed. The analogy is the only thing which should be sought here. And the utilization of mathematics at higher opinions about the natural life should not go any further beyond the aspiration to paralelise the laws of mechanics with their derived laws of relations and quantitative *predictions)* with laws of living nature and its relations. ……. In any use of mathematics for parallellisation one must be aware that the mathematical forms may be used not only as mediating symbols (which is often seen) but that hey must be understood in the full spirit of their meaning. In the same way as for the geometer can use the script with full mathematical correctness and not only as meaningless hieroglyphs. " We can not here comment the whole argumentation, however, it is quite clear that G. de Buquoy understood very well the meaning of mathematical modeling for sciences beyond mechanics, even those which deal with phenomena of life.

This is, in a way which may be directly mapped to modern scientific terminology, has been described already in the book "Sketches form the book of laws of Nature" [4] where the paragraph 185 says:

"We probably may say, that in each actions in the nature there is certain degree of egoism, prevalent subjectivity in the point of action, with exception of one action, movement through mechanical power in non-living bodies."

Which may be directly compared to Caratheódory´s formulation of the second law of thermodynamics [4]

"In any neighbourhood of any state there are states which can not be reached from it by an adiabatic process."

if we know that the only adiabatic process is the mechanical process. Yet, in the "Book of laws of Nature" [4] the generality which is envisioned in "Ideal laudation" [8].has not yet been achieved. And, in fact, it is only now with the advent of econophysics and sociophysics and with the recognition of inavoidability of probabilistic analysis of systems that we are able to recognize it.

The advantage of the Buquoy´s parallelisation approach may be demonstrated in his chemical dynamics [12] and thermodynamics [13]. Since he strictly sticks with parallels with the mechanical theory, he was able to properly recognize some important notions. In chemistry, he correctly recognized existence of "intrinsic rate of chemical reaction" which we now call rate constant, and defined the present notion of reaction rate, quite correctly, as chemical moment. He also in a visionary way describes stoichiometry as "chemical harmony" which may be described as analogous to harmony of music tones.

In thermodynamics Buquoy properly merges the heat and volume expansion effects into one function. This would be perfect match for Gibbs energy if in Buquoys terms this term would not have the character of the moment rather than energy. This approach allowed Buquoy to generally devise heat moment so that it is valid for all materials. The mapping to contemporary, energy-based, thermodynamics, would come through integration of the process trajectory.

In fact, the modern notion of thermodynamic potential has been quite misleading over many years. Namely it brought up the need of a specific mechanical model, which Buquoy always denied. For example in the case of heat radiation he declares: "We have no problem in this case to denote (in agreement with the atomist) this event a luminuous temperature. We do not mean by that a flickering matter (heat push) but its operation line from its central point which nature we do not dare to break in, we do not even hope to." Which is true in the modern sense in two ways: we know that heat is extinguished by heat photons and that interaction of individual particles in any form of matter does not occur purely through a mechanical process – ideally elastic collisions. However, the atomistic model of ideal gas dominates the thermodynamic literature for one and a half century.

# 3 Conclusions

Georg de Buquoy is known to many people. Some celebrate him for creation of the first nature protection area in Central Europe, the Žofín forest. Some recognize his glassworks and the invention of hyalit glass. Others mention his activity in the 1848 revolution. We believe, however, that these reflections are only surface phenomena of the deep recognition of, in modern term, complexity of function of nature. His genial paralellisation approach allowed him to achieve deep insights into laws of nature, insights of more general nature than those of his contemporaries.

A few concrete achievements retained their value till present time, among them the rocket movement equation (2) primarily. Reading Buquoy´s more than 25 books and more than 150 papers always brings new surprises. The main problem of recognition of his achievements resides in language. It may be exemplified in the article [5] which is just a translation of part of much broader description of the generalisation of the Newton law in books published already in 1812 and 1814 [14, 15]. The article [5] is written in French and thus somewhat readable and recognized for present scientific community. Most of Buquoys works are in German, printed in schwabach letters, and thus difficult to read even for part of the contemporary German speaking scientific community.
In any case, reading Buquoy has been extremely interesting and inspirative. By all measures, including the scientometric ones, Buquoy would withstand the criteria of scientist of the beginning of 21$^{st}$ century. The more we read it, the more we agree with Schumpeter that "man and writings are forgotten unjustly, so I think" [3].

**Acknowledgements**

Authors thank, in the order of time, to M. Man for enabling parts of the edition of Buquoys voyage diary, V. Šíma for scientific discussions and the manuscript of his paper [3], for M. von Buquoy and M. K. von Buquoy for articles, technical materials and discussions, to N. Stysova, M.Hokr and V. Bezecny for research and technical collaborations.
This work was supported by CASTECH c.a. and by project „On traces of common history" the Cross-border program Austria-Czech Republic financed by EU 2007-2013.